\documentclass[a4paper,11pt,showkeys]{article}

\usepackage{amsmath,amssymb,amsfonts}
\usepackage{graphicx}
\usepackage{rotating}
\usepackage{rotcapt}
\usepackage[hang,bf,nooneline]{subfigure}
\usepackage[hang,bf,nooneline]{caption}

\newlength{\dslashwidth}

\newcommand{\bsg}{\ensuremath{b\to X_s\gamma}}

\newcommand{\tb}{\ensuremath{\tan\beta}}

\newcommand{\bq}{\begin{equation}}
\newcommand{\eq}{\end{equation}}
\newcommand{\ba}{\begin{array}}
\newcommand{\ea}{\end{array}}
\newcommand{\bqa}{\begin{eqnarray}}
\newcommand{\eqa}{\end{eqnarray}}

\newcommand{\lnf}{{\ifmmode \Lambda^{(N_f)} \else $\Lambda^{(N_f)}$\fi}}
\newcommand{\ms}{{\ifmmode \overline{MS} \else $\overline{MS}$\fi}}
\newcommand{\dr}{{\ifmmode \overline{DR} \else $\overline{DR}$\fi}}
\newcommand{\lms}{{\ifmmode \Lambda^{(5)}_{\overline{MS}} \else $\Lambda^{(5)}_{\overline{MS}}$\fi}}
\newcommand{\lam}{{\ifmmode \Lambda \else $\Lambda$\fi}}
\newcommand{\gev}{{\ifmmode {\rm GeV} \else ${\rm GeV}$\fi}}
\newcommand{\gevc}{{\ifmmode {\rm GeV/c^2} \else ${\rm GeV/c^2}$\fi}}
\newcommand{\tev}{{\ifmmode {\rm TeV} \else ${\rm TeV}$\fi}}
\newcommand{\tevc}{{\ifmmode {\rm TeV/c^2} \else ${\rm TeV/c^2}$\fi}}
\newcommand{\lp}{{\ifmmode L^+  \else $L^+$\fi}}
\newcommand{\lm}{{\ifmmode L^-  \else $L^-$\fi}}
\newcommand{\mlp}{{\ifmmode M(L^-) \else $M(L^-)$\fi}}
\newcommand{\mlz}{{\ifmmode M(L^0) \else $M(L^0)$\fi}}
\newcommand{\lz}{{\ifmmode L^0 \else $L^0$\fi}}
\newcommand{\ev}{{\ifmmode GeV/c^2 \else $GeV/c^2$\fi}}
\newcommand{\tri}{{\ifmmode \triangleup \else $\triangleup$\fi}}
\newcommand{\unl}{{\ifmmode U_{lL^0} \else $U_{lL^0}$\fi}}\newcommand{\gL}{{\ifmmode g_L \else $g_{L}$\fi}}
\newcommand{\gR}{{\ifmmode g_R  \else $g_{R}$\fi}}
\newcommand{\gumu}{{\ifmmode \gamma^{\mu} \else $\gamma^{\mu}$\fi}}
\newcommand{\gunu}{{\ifmmode \gamma^{\nu} \else $\gamma^{\nu}$\fi}}
\newcommand{\gdmu}{{\ifmmode \gamma_{\mu} \else $\gamma_{\mu}$\fi}}
\newcommand{\gdnu}{{\ifmmode \gamma_{\nu} \else $\gamma_{\nu}$\fi}}
\newcommand{\stw}{{\ifmmode\sin^2\theta_W \else $\sin^{2}\theta_{W}$ \fi}}
\newcommand{\sws}{{\ifmmode \;\sin^2\theta_W  \else $\;\sin^{2}\theta_{W}$ \fi}}
\newcommand{\cws}{{\ifmmode \;\cos^2\theta_W  \else $\;\cos^{2}\theta_{W}$ \fi}}
\newcommand{\sw}{{\ifmmode \;\sin\theta_W  \else $\sin\theta_{W}$ \fi}}
\newcommand{\cw}{{\ifmmode \;\cos\theta_W  \else $\;\cos\theta_{W}$ \fi}}
\newcommand{\tw}{{\ifmmode \;\tan\theta_W  \else $\;\tan\theta_{W}$ \fi}}
\newcommand{\qq}{{\ifmmode q\overline{q} \else $q\overline{q}$\fi}}
\newcommand{\lR}{{\ifmmode l_R \else $l_R$\fi}}
\newcommand{\lL}{{\ifmmode l_L \else $l_L$\fi}}
\newcommand{\nt}{{\ifmmode \nu_{\tau} \else $\nu_{\tau}$\fi}}
\newcommand{\nuR}{{\ifmmode \nu_R  \else $\nu_R$\fi}}
\newcommand{\nuL}{{\ifmmode \nu_L  \else $\nu_L$\fi}}
\newcommand{\qR}{{\ifmmode g_R  \else $q_R$\fi}}
\newcommand{\qL}{{\ifmmode q_L  \else $q_L$\fi}}
\newcommand{\qRp}{{\ifmmode q_R'  \else $q_{R}$'\fi}}
\newcommand{\qLp}{{\ifmmode q_L'  \else $q_{L}$'\fi}}
\newcommand{\est}{{\ifmmode e^{\bf \ast} \else $e^{\bf \ast}$\fi}}
\newcommand{\lst}{{\ifmmode l^{\bf \ast} \else $l^{\bf \ast}$\fi}}
\newcommand{\must}{{\ifmmode \mu^{\bf \ast} \else $\mu^{\bf \ast}$\fi}}
\newcommand{\taust}{{\ifmmode \tau^{\bf \ast} \else $\tau^{\bf \ast}$ \fi}}
\newcommand{\pperp}{{\ifmmode p_t  \else $p_t$\fi}}
\newcommand{\et}{{\ifmmode E_t  \else $E_t$\fi}}
\newcommand{\xt}{{\ifmmode x_t  \else $x_t$\fi}}
\newcommand{\smumu}{{\ifmmode \sigma_{\mu\mu}  \else $\sigma_{\mu\mu}$ \fi}}
\newcommand{\eg}{{\ifmmode e\gamma  \else $e\gamma$\fi}}
\newcommand{\epem}{{\ifmmode e^+e^-  \else $e^+e^-$\fi}}
\newcommand{\lplm}{{\ifmmode L^+L^-  \else $L^+L^-$\fi}}
\newcommand{\pp}{{\ifmmode p\overline p  \else $p\overline p$\fi}}
\newcommand{\llz}{{\ifmmode L^0\overline{L}^0 \else $L^0\overline{L}^0$\fi}}
\newcommand{\epemt}{{\ifmmode e^+e^- \to  \else $e^+e^- \to$\fi}}
\newcommand{\eb}{{\ifmmode E_{beam}  \else $E_{beam}$\fi}}
\newcommand{\ip}{{\ifmmode pb^{-1}  \else $pb^{-1}$\fi}}
\newcommand{\upm}{{\ifmmode ^{\pm}  \else $^{\pm}$\fi}}
\newcommand{\de}{{\ifmmode ^{\circ}  \else $^{\circ}$ \fi}}
\newcommand{\appr}{{\ifmmode \sim \else $\sim$ \fi}}
\newcommand{\corresp}{{\ifmmode \stackrel{\wedge}{=} \else $\stackrel{\wedge}{=}$ \fi}}
\newcommand{\sqrts}{{\ifmmode \sqrt{s} \else $\sqrt{s}$\fi}}
\newcommand{\zz}{{\ifmmode Z^0  \else $Z^0$\fi}}
\newcommand{\mz}{{\ifmmode M_{Z}  \else $M_{Z}$\fi}}
\newcommand{\mzs}{{\ifmmode M_{Z}^2  \else $M_{Z}^2$\fi}}
\newcommand{\mw}{{\ifmmode M_{W}  \else $M_{W}$\fi}}
\newcommand{\mws}{{\ifmmode M_{W}^2  \else $M_{W}^2$\fi}}
\newcommand{\mh}{{\ifmmode M_{Higgs}  \else $M_{Higgs}$\fi}}
\newcommand{\gt}{{\ifmmode \Gamma_{tot} \else $\Gamma_{tot}$\fi}}
\newcommand{\msusy}{{\ifmmode M_{SUSY}  \else $M_{SUSY}$\fi}}
\newcommand{\msusys}{{\ifmmode M_{SUSY}^2  \else $M_{SUSY}^2$\fi}}
\newcommand{\su}{{\ifmmode SU(3)_C\otimes\- SU(2)_L\otimes\- U(1)_Y \else $SU(3)_C\otimes SU(2)_L\otimes U(1)_Y$\fi}}
\newcommand{\suthree}{{\ifmmode SU(3)_C  \else $SU(3)_C$\fi}}
\newcommand{\sutwo}{{\ifmmode  SU(2)_L\otimes U(1)_Y \else $SU(2)_L\otimes U(1)_Y$\fi}}
\newcommand{\taup} {{\ifmmode \tau_{proton} \else $\tau_{proton}$\fi}}
\newcommand{\as}{{\ifmmode \alpha_{s}  \else $\alpha_{s}$\fi}}
\newcommand{\mgut}{{\ifmmode M_{GUT}  \else $M_{GUT}$\fi}}
\newcommand{\mguts}{{\ifmmode M_{GUT}^2  \else $M_{GUT}^2$\fi}}
\newcommand{\mze} {{\ifmmode m_0        \else $m_0$\fi}}
\newcommand{\mha}{{\ifmmode m_{1/2}    \else $m_{1/2}$\fi}}
\newcommand{\mb} {{\ifmmode m_{b}    \else $m_{b}$\fi}}
\newcommand{\mt} {{\ifmmode m_{t}    \else $m_{t}$\fi}}
\newcommand{\mts} {{\ifmmode m_{t}^2    \else $m_{t}^2$\fi}}

\newcommand{\mtau}{{\ifmmode m_{\tau}  \else $m_{\tau}$\fi}}
\newcommand{\dpp}{{\ifmmode \delta_{pert} \else $\delta_{pert}$\fi}}
\newcommand{\dnp}{{\ifmmode\delta_{non-pert}\else$\delta_{non-pert}$\fi}}
\newcommand{\dew}{{\ifmmode \delta_{\rm EW}\else $\delta_{\rm EW}$\fi}}
\newcommand{\rt}{{\ifmmode R_{\tau}  \else $R_{\tau} $\fi}}
\newcommand{\rz}{{\ifmmode R_{Z}  \else $R_{Z} $\fi}}

\newcommand{\swb}{{\ifmmode \sin^2\theta_{\overline{MS}} \else $\sin^2\theta_{\overline{MS}}$\fi}}
\newcommand{\cwb}{{\ifmmode \cos^2\theta_{\overline{MS}} \else $\cos^2\theta_{\overline{MS}}$\fi}}

\newcommand{\mzero}{{\ifmmode m_0 \else $m_0$\fi}}
\newcommand{\mhalf}{{\ifmmode m_{1/2} \else $m_{1/2}$\fi}}

\begin{document}

\begin{center}
\Large \textbf{The supersymmetric interpretation of the \\EGRET
excess of diffuse Galactic gamma rays}

\vspace{10mm}

\large

W. de Boer$^{1}$, C. Sander$^{1}$, V. Zhukov$^{1}$,\\
A.V. Gladyshev$^{2,3}$, D.I. Kazakov$^{2,3}$

\normalsize \vspace{5mm}
$^1$ \textit{Institut f\"ur Experimentelle Kernphysik, Universit\"at Karlsruhe (TH),\\
P.O. Box 6980, 76128 Karlsruhe, Germany}

\vspace{5mm}
$^2$ \textit{Bogoliubov Laboratory of Theoretical Physics, Joint Institute for Nuclear Research,\\
141980, 6 Joliot-Curie, Dubna, Moscow Region, Russia}

\vspace{5mm}
$^3$ \textit{Institute for Theoretical and Experimental Physics,\\
117218, 25 B.Cheremushkinskaya, Moscow, Russia}

\vspace{30mm} \textbf{Abstract} \vspace{5mm}

\begin{minipage}[c]{12cm}

\textit{Recently it was shown that the excess of diffuse Galactic
gamma rays above 1 GeV traces the Dark Matter halo, as proven by
reconstructing the peculiar shape of the  rotation curve of our
Galaxy from the gamma ray excess. This can be interpreted as a Dark
Matter annihilation signal. In this paper we investigate if this
interpretation is consistent with Supersymmetry. It is found that
the EGRET excess combined with  all electroweak constraints  is
fully consistent with the minimal mSUGRA model for scalars in the
TeV range and gauginos below 500 GeV.}

\end{minipage}
\end{center}

\thispagestyle{empty} \setcounter{page}{0} \clearpage

\section{Introduction}

Cold Dark Matter (CDM) makes up 23\% of the energy of the universe,
as deduced from the temperature anisotropies in the Cosmic Microwave
Background (CMB) in combination with data on the Hubble expansion
and the density fluctuations in the universe~\cite{wmap}. One of the
most popular CDM candidates is the neutralino, a stable neutral
particle predicted by Supersymmetry~\cite{lspdm,jungman}. The
neutralinos are spin 1/2 Majorana particles, which can annihilate
into pairs of Standard Model (SM) particles. A large fraction of the
annihilations is expected to go into quark-antiquark pairs. Since
the DM particles are strongly non-relativistic, the initial energy
is simply given by two times the neutralino mass, which is converted
into energy of the quarks, which are then mono-energetic. In a
recent paper we showed that the observed excess of diffuse Galactic
gamma rays has all the properties of the $\pi^0$ decays of such
mono-energetic quarks originating from the annihilation of
neutralinos with a mass around 60 GeV \cite{us,sander}. For a better
understanding of the following we shortly summarize these results.

Gamma rays from Dark Matter Annihilation (DMA) can be distinguished
from the background (BG) by their completely different spectral
shape: the background originates mainly from cosmic rays (CR)
hitting the gas of the disc and producing abundantly $\pi^0$ mesons,
which decay into two photons. The initial CR spectrum is a steep
power law spectrum, which yields a much softer gamma ray spectrum
than the fragmentation of the hard mono-energetic quarks from DMA.
The spectral shape of the gamma rays from the background is well
known from fixed target experiments given the known CR spectrum. The
spectral shape of the gamma rays from DMA is well known from the
fragmentation of mono-energetic quarks studied at electron-positron
colliders, like LEP at CERN, which has been operating up to
centre-of-mass energies of about 200 GeV, i.e. it corresponds to
gamma spectra from neutralino masses up to 100 GeV. The different
quark flavours all yield similar gamma spectra at high energies. In
addition to these two main components with a shape well known from
accelerator experiments, there are contributions from inverse
Compton scattering and Bremsstrahlung. In the gamma ray energy range
of interest (above 0.1 GeV) these contributions are small, but their
shape is well known too.

Experimentally, the spectral shape of the diffuse Galactic gamma
rays has been  measured with the EGRET satellite in the range 0.1 to
10 GeV. The EGRET data are publicly available as high resolution
(0.5$^\circ$) sky maps from the NASA archive\footnote{NASA archive:
\texttt{http://cossc.gsfc.nasa.gov/archive/.}}, which allows an
independent analysis in many different sky directions\cite{us}.
Comparing the BG with the EGRET data shows that above 1 GeV there is
a large deficit of gamma rays, which reaches more than a factor of
two towards the Galactic centre. However, fitting two components,
namely BG and DMA, yields a perfect fit in all sky directions for a
DM particle mass around 60 GeV. From the normalization factors for
the BG and DMA components in 180 independent sky directions the
distribution of DM has been obtained. Combining this with the known
distribution of the visible matter yields the complete mass
distribution, which in turn can be used to reconstruct the rotation
curve of our Galaxy. The surprise was, that the gamma rays indeed
explain the peculiar structure of this rotation curve, which was
found to originate from substructure in the DM halo.

So the famous EGRET excess of diffuse Galactic gamma rays, discussed
already in 1997 \cite{hunter}, was found to possess all the expected
properties from DMA: it is observable in all sky directions and has
everywhere the shape expected for the annihilation of DM particles
with a mass around 60 GeV. In addition, the reconstruction of the
rotation curve from the EGRET excess proves that the latter traces
the DM in our Galaxy \cite{us}. Note that the evidence for DMA is
not so much the amount of excess, but how it is distributed in the
sky and that it has the same spectral shape in all sky directions.
Any unknown systematic errors in the EGRET data are expected to be
independent of the sky direction, so the evidence for DMA does not
depend on ``unknown unknowns''.

It is the purpose of the present paper to see if this intriguing
hint of DMA is compatible with Supersymmetry. Here we will
concentrate on the Minimal Supersymmetric Model with supergravity
inspired symmetry breaking (mSUGRA model)\cite{susyrev}. We assume
that the EGRET excess originates from the annihilation of the
stable, neutral lightest supersymmetric particles, the neutralinos.
Their mass is then constrained to be between 50 and 100 GeV from the
EGRET data, which strongly constrains the masses from all other SUSY
particles, if mass unification at the GUT scale is assumed. It will
be shown that combining the EGRET data with other constraints, like
the electroweak precision data, Higgs mass limits, chargino limits,
radiative electroweak symmetry breaking and relic density leads to a
very constrained SUSY mass spectrum with light gauginos and heavy
squarks and sleptons.

\section{Comparison  with mSUGRA} \label{susywimp}

The mSUGRA model, i.e. the Minimal Supersymmetric Standard Model
(MSSM) with supergravity inspired breaking terms, is characterised
by only 5 parameters: $\mzero$, $\mhalf$, $\tb$, sign($\mu$), $A_0$
\cite{susyrev}. Here $m_0$ and $\mhalf$ are the common masses for
the gauginos and scalars at the GUT scale. The latter is determined
by the unification of the gauge couplings. Gauge unification is
still possible with the precisely measured couplings at
LEP~\cite{bs}. The ratio of the vacuum expectation values of the
neutral components of the two Higgs doublets in Supersymmetry is
called \tb ~ and $A_0$ is the trilinear coupling at the GUT scale.
We only consider the dominant trilinear couplings of the third
generation of quarks and leptons and assume also $A_0$ to be unified
at the GUT scale. Electroweak symmetry breaking fixes the scale of
$\mu$ \cite{susyrev}, so only its sign is a free parameter. We use
the positive sign, as suggested by the small deviation of the muon
anomalous moment from the Standard Model (SM) \cite{bs}.

\begin{figure}
 \begin{center}
  \includegraphics [width=0.6\textwidth,clip]{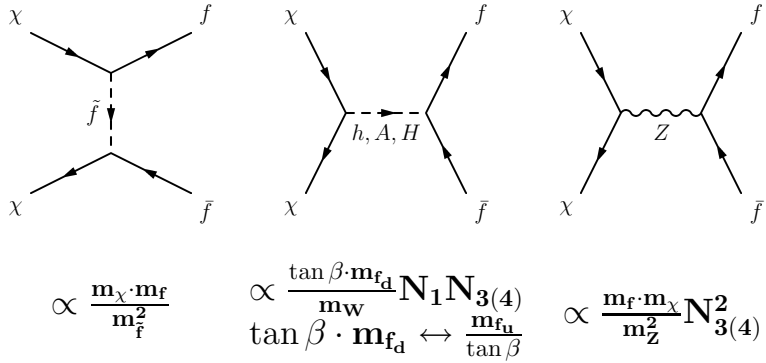}
  \caption[]{The dominant annihilation diagrams for the lightest neutralino, which is a
  linear combination of the gaugino and Higgsino states: $|\chi_0\rangle =N_1 |B_0\rangle +N_2|W^3_0\rangle +N_3|H_1\rangle +N_4|H_2\rangle$. The dependence of the amplitudes on masses and neutralino mixing parameter $N_i$ has been indicated.} \label{diagrams}
 \end{center}
\end{figure}

The dominant annihilation diagrams of the lightest supersymmetric
particle (LSP) neutralino are shown in figure \ref{diagrams}. The
cross sections are proportional to the final state fermion mass,
which originates either from the Yukawa couplings for the Higgs
exchange diagram or from the helicity suppression at the low
energies involved in cold DMA \cite{goldberg}. Therefore heavy
fermion final states, i.e. third generation quarks and leptons, are
expected to be dominant. The Higgs exchange diagram is in addition
proportional to \tb ~ for down type quarks and 1/\tb ~ for up type
quarks, indicating that top quark final states are suppressed for
large \tb. The W- and Z-final states from t-channel chargino and
neutralino exchange have usually a much smaller cross section due to
the weak couplings involved and are in addition kinematically
suppressed for the 60 GeV neutralino mass preferred by the EGRET
data.

\begin{figure}
 \begin{center}
  {\includegraphics[width=0.45\textwidth]{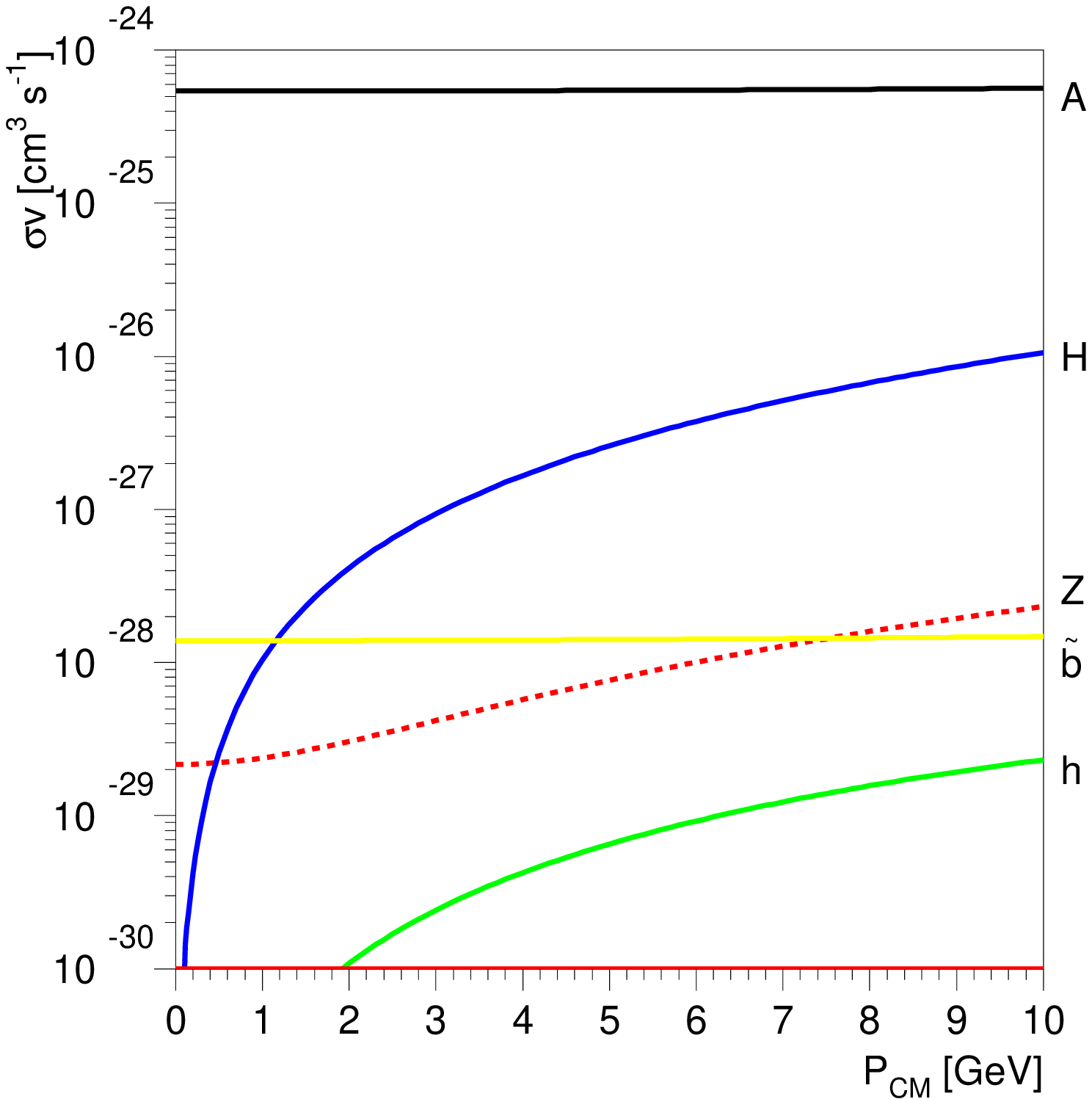}}
  {\includegraphics[width=0.45\textwidth]{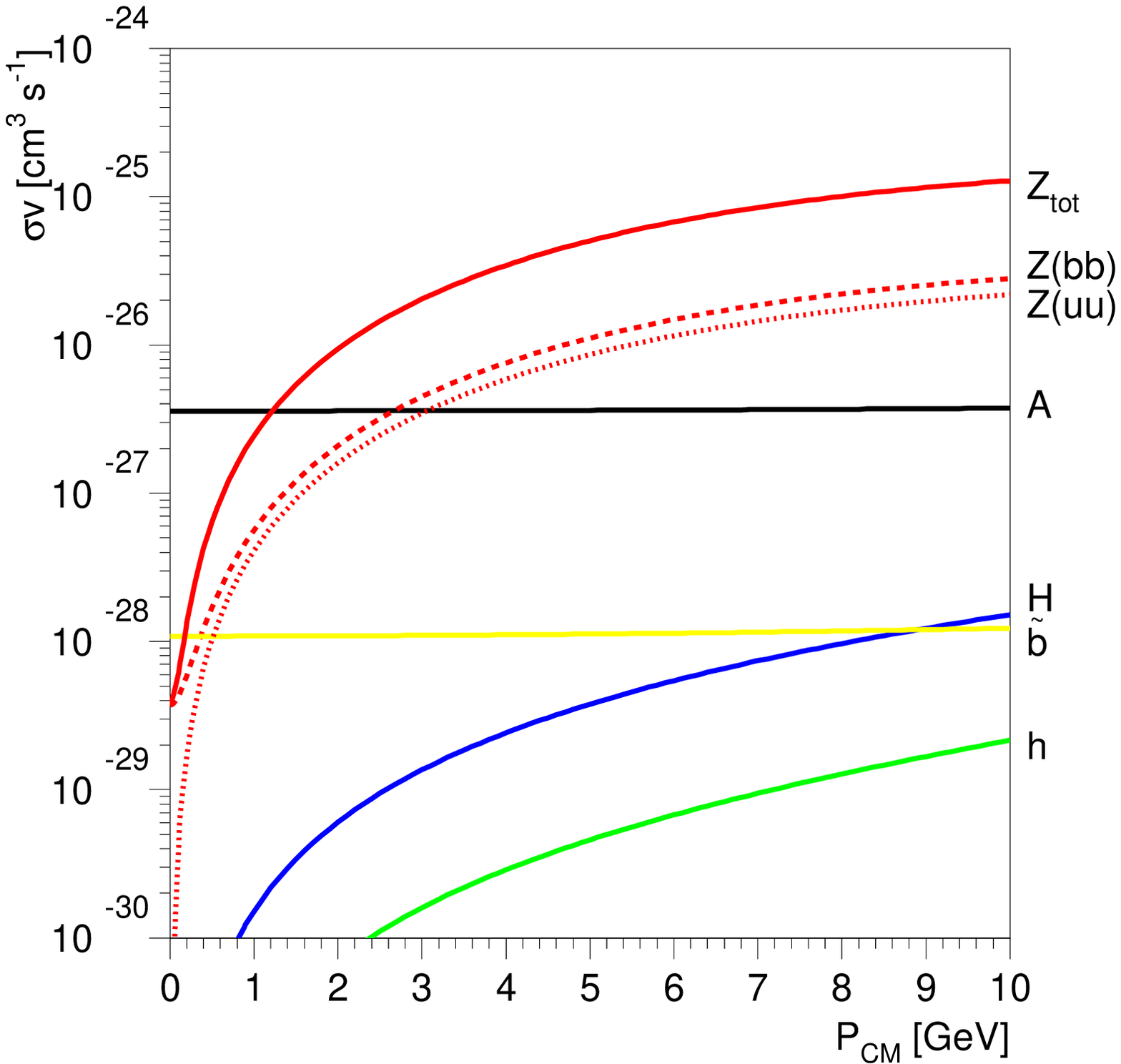}}
  \caption[]{The neutralino annihilation cross section $\sigma$ multiplied by the relative neutralino velocities ($v$) for the different diagrams of Fig. \ref{diagrams} (indicated by the exchanged particle) as function of the center of mass momentum for two different sets of parameters: left: the annihilation via the pseudoscalar Higgs $A$ is dominant (heavy scalars, large \tb); right: the dominant channel is through the $Z$-Boson ($m_\chi\sim m_Z/2$).} \label{sigmap}
 \end{center}
\end{figure}

The annihilation rate, which is proportional to the cross section
multiplied by the relative neutralino velocities, is practically
independent of the centre of mass energy for the pseudoscalar Higgs
and sfermion exchange diagrams, but strongly dependent on energy for
the other diagrams, as was calculated with the CalcHEP package
\cite{calchep} and shown in Fig. \ref{sigmap}. This implies that for
the present temperature of the universe close to absolute zero the
neutralino annihilation is dominated by either sfermion exchange or
pseudoscalar Higgs exchange. The sfermion exchange is suppressed for
the following reason. The Born mass of the lightest Higgs is below
the $Z^0$ mass, but radiative corrections can boost it up to 130 GeV
in the minimal mSUGRA model\cite{susyrev}. These corrections depend
on the heavy particles coupling to the lightest Higgs, like the top
and stop quarks.  These scalars have to be sufficiently heavy in
order to reach a Higgs mass above 114 GeV, which is the present
lower limit from the direct searches at LEP \cite{lephiggs}. The
Higgs mass was calculated with the FeynHiggs program
\cite{feynhiggs}. Note that this Higgs limit from LEP is the limit
on the Standard Model Higgs particle, but for heavy scalars the
lightest SUSY Higgs particle has very much the properties of the SM
Higgs, so the above limit is also valid in our case, as will be
shown below.

\begin{figure}
 \begin{center}
  \includegraphics [width=0.45\textwidth,clip]{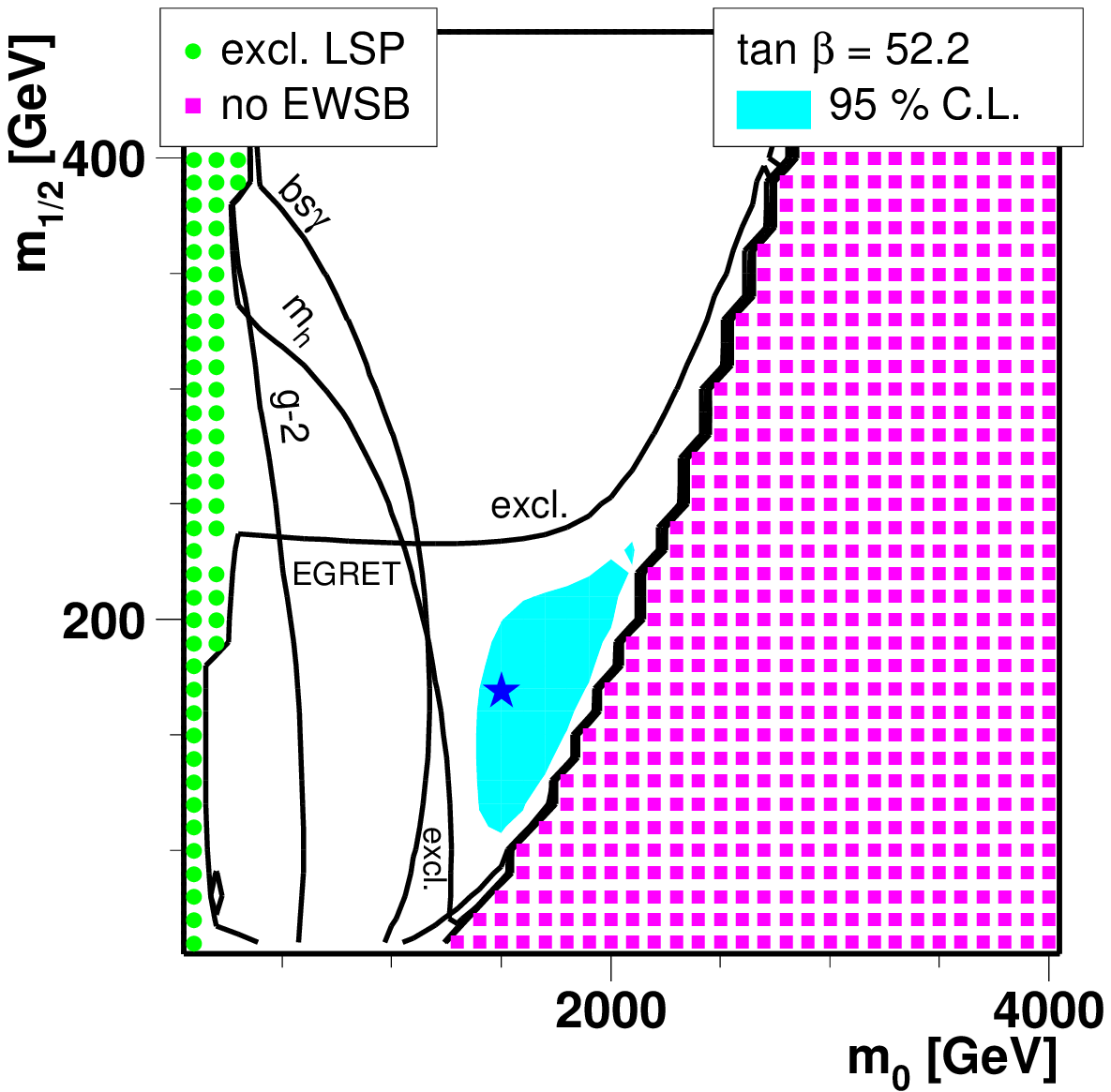}
  \includegraphics [width=0.45\textwidth,clip]{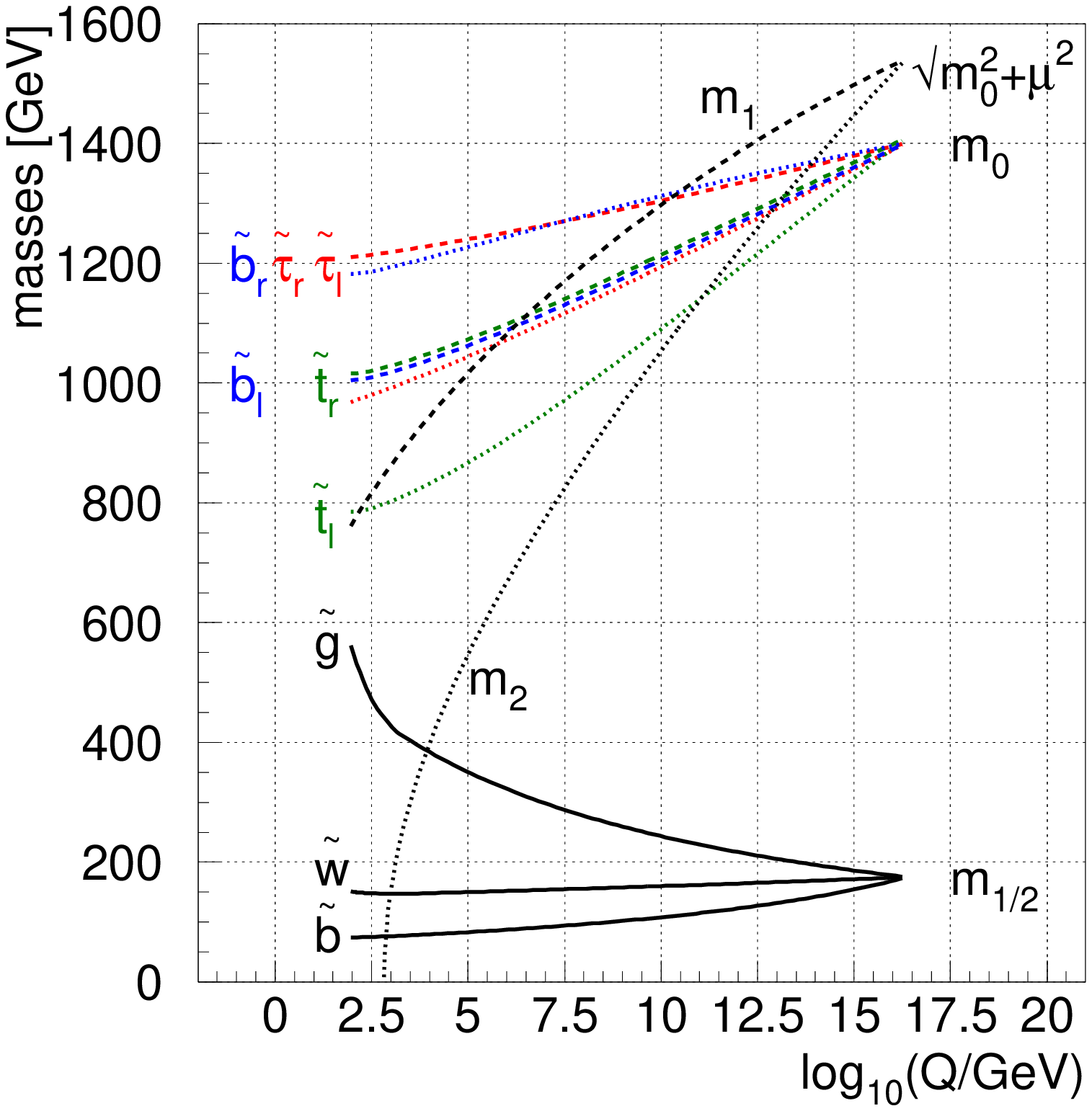}
  \caption[]{Left: the light shaded area (blue) indicates the 95\% C.L. parameter
  range in the $\mzero$-$\mhalf$-plane allowed by the EGRET data, if the constraints
  from electroweak data, a neutral LSP and electroweak symmetry breaking (EWSB) are imposed as well.
  The individual constraints have been indicated by the lines and dots.
  For the left hand panel the values of $A_0=0$ and $\tb=52.2$ were
  chosen and the choice of parameters of Table \ref{t1} has been
  indicated by a star.
  The right hand side shows the running SUSY masses for a set of allowed parameters.} \label{msugra}
 \end{center}
\end{figure}

 For light neutralinos, i.e. small $\mhalf$, the Higgs mass limit requires
$\mzero$ to be in the TeV range, as indicated in the left hand panel
of Fig. \ref{msugra} by the almost vertical line, labeled $m_h$. The
EGRET data requires $\mhalf$ to be below the almost horizontal line
at $\mhalf=230$ GeV. The low value of $\mhalf$ implies that the
gluino QCD corrections to the stop quarks are rather small, so the
``bare'' quark mass $\mzero$ has to be large to obey the Higgs
limit, which requires heavy stops. In addition, the excluded regions
from $\bsg$ and the anomalous magnetic moment of the muon have been
indicated (left from the corresponding lines). These latter values
were calculated with the publicly available web-based program from
Ref. \cite{kraml}, which uses \texttt{micrOMEGAs 1.4}
\cite{micromegas} for the relic density calculation and we opted for
the SUSY mass spectrum from the \texttt{Suspect 2.3.4} program
\cite{suspect}. For the exclusion limits the following inputs were
used: a) $Br(B \to X_s\gamma)=(3.43\pm 0.36)\cdot 10^{-4}$, which is
the average from BaBar, CLEO and BELLE \cite{bsgexp} and b) the
deviation of the anomalous magnetic moment of the muon $a_\mu$ from
the expected value in the Standard Model was taken to be \cite{amu}:
$\Delta
a_\mu=a_\mu^{\mbox{\scriptsize{exp}}}-a_\mu^{\mbox{\scriptsize{theo}}}=(27\pm
10)\cdot 10^{-10}$.

The lower limits on $\mzero$ discussed above are practically
independent of $A_0$ due to a coincidence from the constraints from
the \bsg~ rate and the lower limit on the Higgs mass of 114
GeV~\cite{bs}. The absolute value of the Higgs mixing parameter
$\mu$ is determined by electroweak symmetry breaking, while its sign
is taken to be positive, as preferred by the anomalous magnetic
moment of the muon. The region of large $\mzero$, for which no
electroweak symmetry breaking (EWSB) is possible, has been indicated
in Fig. \ref{msugra} as well as the region of small $\mzero$, where
the stau would be the lightest SUSY particle, which is excluded,
since the DM candidate has to be neutral.

The $\mzero$ values in the TeV range between the Higgs mass limit
and EWSB limit are allowed by all constraints considered sofar. It
should be noted that these boundaries are quite sensitive to the
gauge and Yukawa couplings, which determine the radiative
corrections to the Higgs mass and the radiative corrections to the
Higgs potential needed for EWSB. E.g. increasing the top mass from
175 to 178 GeV increases the lightest Higgs mass by about 1 GeV,
which can be compensated by lowering $\mzero$ by 200 GeV. Thus the
error on the curve labeled $m_h$ is around 200 GeV in the horizontal
direction in the range below the EGRET line. The uncertainty on the
EWSB region is even larger at large values of $\tb$. Increasing the
top mass from 175 to 178 GeV moves the EWSB boundary by
approximately 1 TeV to the right. This sensitivity can be understood
as follows. Electroweak symmetry breaking is triggered if the
following condition is fulfilled:
\begin{equation}
  {M_Z^2\over 2} = {m_1^2 - m_2^2\tan^2\beta \over \tan^2\beta-1} \approx -m_2^2. \label{ewsb}
\end{equation}
Here $m_1$ and $m_2$ are the mass parameters in the Higgs potential
with two Higgs doublets\cite{susyrev}. The last term is valid for
large \tb, implying $m_2$ becoming negative for a positive value of
$M_Z^2$ and large $\tb$ is required by the relic density, as will be
discussed below. EWSB is then possible for large values of the top
Yukawa coupling, which drives $m_2$ negative, as shown on the right
hand side of Fig. \ref{msugra}, where the running masses
corresponding to a set of GUT scale parameters compatible with all
constraints are shown. The starting value of $m_2$ at the GUT scale
is $\sqrt{m_0^2+\mu^2}$. The running of $m_2$ over 14 orders of
magnitude between the GUT scale and the electroweak scale implies
that a small increase in the top Yukawa coupling, thus increasing
the slope of the running, requires a large increase in $m_0$.

In summary: increasing the top mass by 3 GeV widens the allowed
region in Fig. \ref{msugra}: 200 GeV to lower values of $m_0$ and
around 1 TeV to larger values of $m_0$, but from the running of the
masses in Fig. \ref{msugra} it is clear that the low energy masses
for squarks and sleptons are in the TeV range or above and the
gauginos below 500 GeV. The  spectrum and the corresponding values
of the relic density $\Omega h^2$, $\bsg$ and $\Delta a_\mu$ have
been tabulated in Table \ref{t1} for a typical set of parameters
compatible with all constraints.

\begin{figure}
 \begin{center}
  \includegraphics [width=0.52\textwidth,clip]{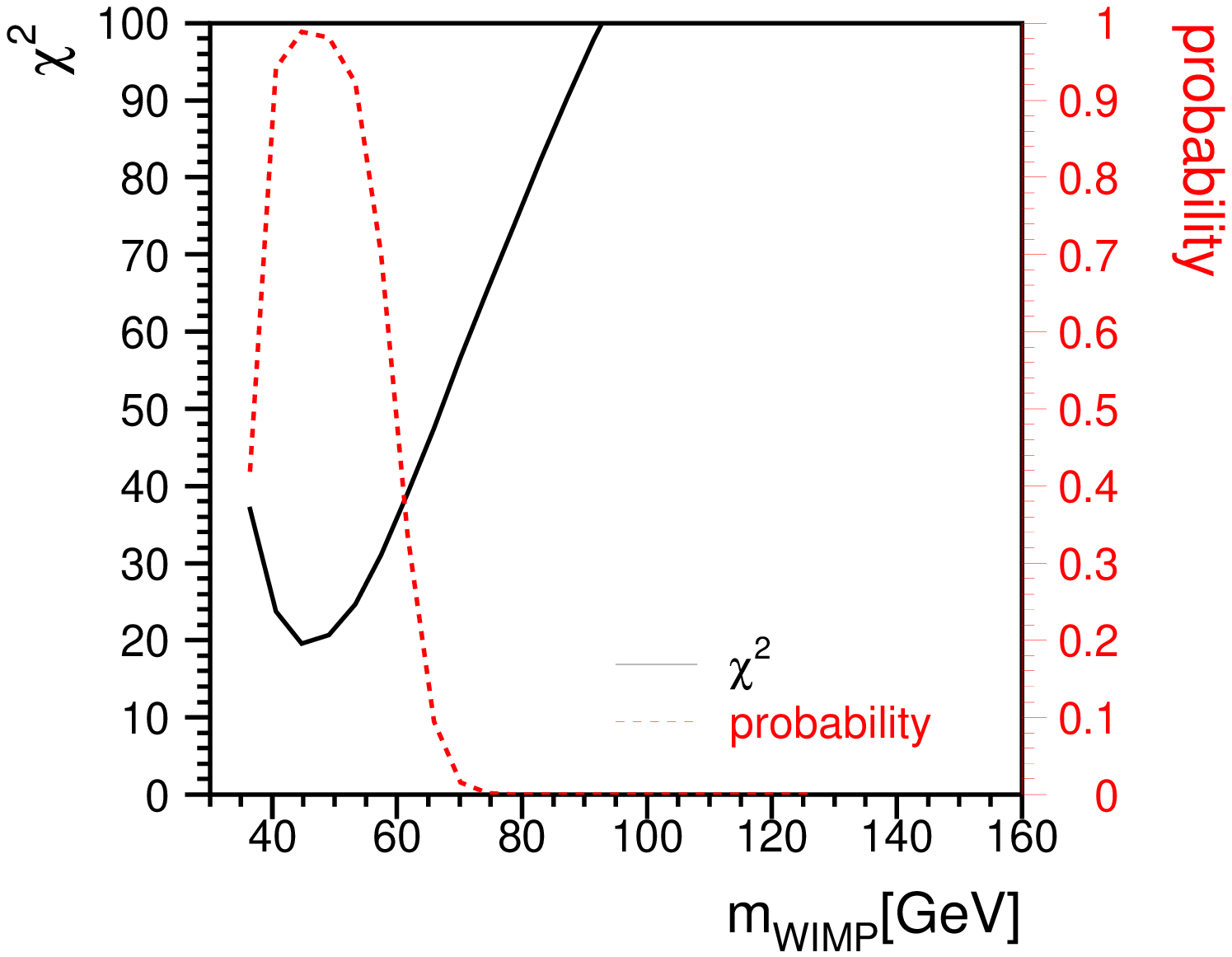}
  \includegraphics [width=0.39\textwidth,clip]{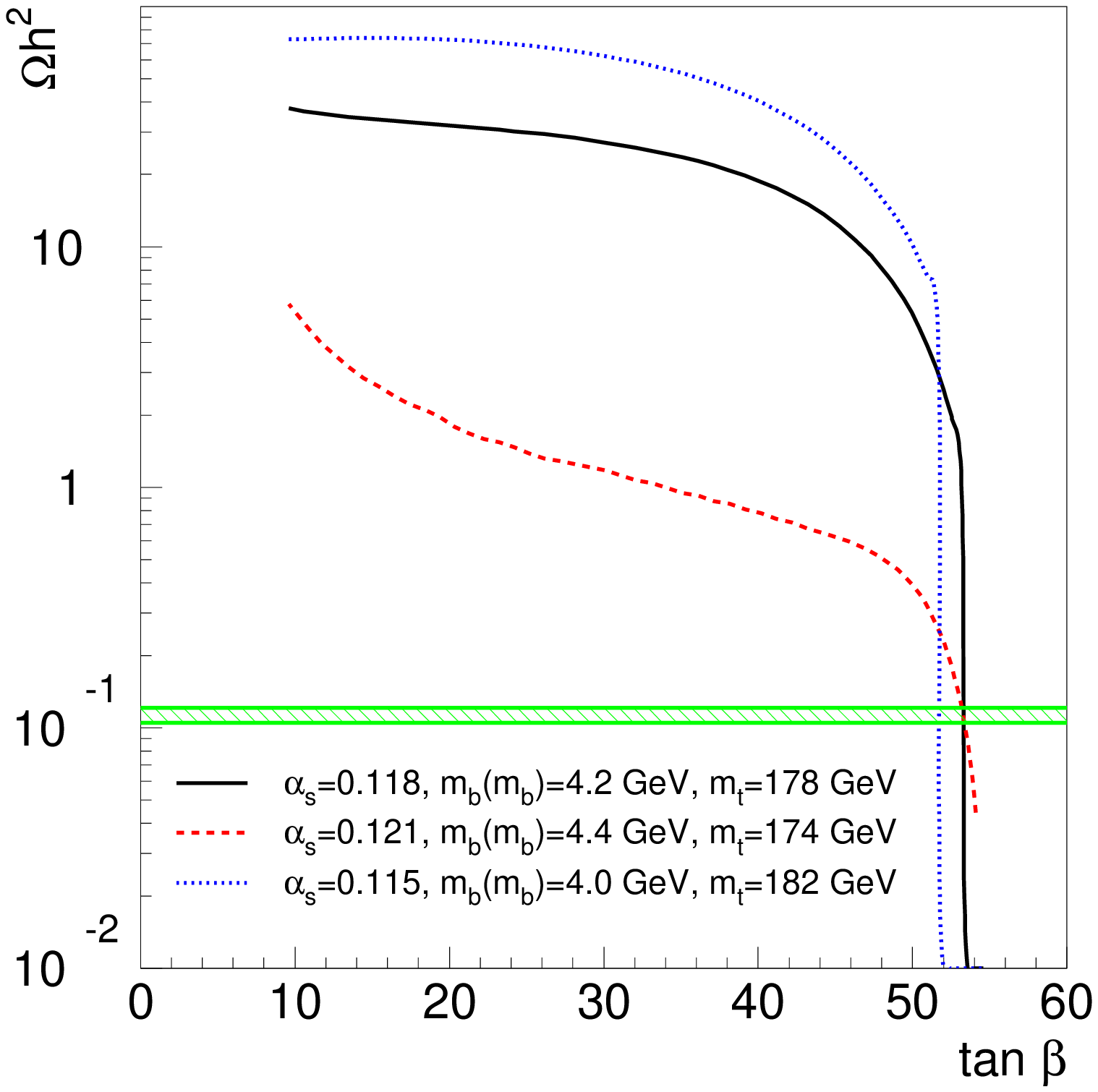}
  \caption[]{Left hand side:  the $\chi^2$ distribution and corresponding probability as function of
  the WIMP mass from a fit to the EGRET data on galactic gamma rays.
 The 95\% CL  upper limit on the WIMP mass is 70 GeV for the conventional background model.
 The limit can be stretched to 100 GeV for a model maximizing the background \cite{sander}.
 On the right hand side $\Omega h^2$ as a function of $\tb$ is plotted for $\mzero =1500$ GeV,
 $\mhalf =200$ GeV, $A_0=0$.  The horizontal shaded band corresponds to
the observed relic density.} \label{uncertainties}
 \end{center}
\end{figure}

As discussed above the uncertainty in $m_0$ is large, but the 95\%
C.L. upper limit on the lightest neutralino is around 70 GeV, as can
be deduced from the $\chi^2$ fit to the EGRET energy spectrum of the
diffuse gamma rays. This $\chi^2$ distribution is shown in Fig.
\ref{uncertainties} together with the probabilities. Above 70 GeV
the fit probability is below 5\%, implying $\mhalf <175$ GeV.
However, for background models, which try to maximize the background
by assuming that the local cosmic ray spectrum is not representative
for our galaxy,  neutralino values up to 100 GeV can be obtained
\cite{us,sander}, as indicated in Fig. \ref{msugra} by the ``EGRET''
line. A lower limit of 50 GeV on the neutralino mass avoids the
dominance of the $Z^0$ cross section in the annihilation, since if
the neutralino mass is close to half of the $Z^0$ mass the
annihilation signal at the present temperature of the universe would
be practically zero, as demonstrated in Fig. \ref{sigmap} in
disagreement with the EGRET excess (unless one allows
unrealistically large clumping of the DM, which increases the
annihilation rate). The lightest neutralino is a mixture of all spin
1/2 neutral particles: $|\chi_o\rangle =N_1|B_0\rangle
+N_2|W^3_0\rangle +N_3|H_1\rangle +N_4|H_2\rangle$ with
$(N_1,N_2,N_3,N_4)=(0.95,-0.10,0.27,-0.09)$ for the values of Table
\ref{t1} meaning that the lightest neutralino is an almost pure bino
for the allowed region of Fig. \ref{msugra}.

The correct value of the relic density is obtained for large \tb, as
shown on the right hand side of Fig. \ref{uncertainties} for a given
set of SUSY mass parameters and different values of the SM
parameters $\alpha_s$, $m_t$ and $m_b$. The relic density was
calculated with \texttt{micrOMEGAs 1.4} \cite{micromegas}. Scanning
over the allowed region of Fig. \ref{msugra} {\it and} requiring an
LSP mass above 50 GeV requires $\tb$ to be in the range of 50 to 55.
Note that the EGRET data itself is only sensitive to the masses, not
to \tb. But for practically any set of masses the correct relic
density can be obtained by a suitable value of \tb~ and $A_0$
\cite{sander}. The strong dependence of the relic density on \tb~
for large values of \tb~ originates from the strong dependence of
the pseudoscalar Higgs mass on \tb, which  in mSUGRA is given by:
\begin{equation}
  m_A^2=m_1^2+m_2^2=m_{H_1}^2+m_{H_2}^2+2\mu^2,
\end{equation}
where  $m_{H_i}$ are the Higgs mass terms.
 However, at large \tb~ $m_1^2$ is
also driven negative, since then the bottom Yukawa coupling becomes
of the same order as magnitude as the top Yukawa coupling. This can
be seen as follows: the top and bottom masses are given by:
\begin{eqnarray}
  m_t^2 &=& h_t^2 \cdot\vert H_2\vert^2 = h_t^2 \cdot v_2^2 = h_t^2 \cdot |v|^2 \sin^2\beta \nonumber \\
  m_b^2 &=& h_b^2 \cdot\vert H_1\vert^2 = h_b^2 \cdot v_1^2= h_b^2 \cdot |v|^2 \cos^2\beta,
\end{eqnarray}
so if the ratio $ \tb\approx 50 \approx m_t/m_b$ then the Yukawa
couplings $h_t$ and $h_b$ must be of the same order of magnitude. If
both $m_1$ and $m_2$ are driven negative, the pseudoscalar Higgs
mass can only become positive by large radiative corrections from
stop and sbottom quarks, which works if the latter are heavy. But
large radiative corrections lead to a large variation in the
pseudoscalar Higgs mass and a correspondingly large variation in the
relic density, as shown in Fig. \ref{uncertainties}. Note that it is
interesting that the EGRET scenario combined with the relic density
from WMAP requires $\tb$ to be in the range, where the Yukawa
couplings of top, bottom and tau can be unified\cite{zp}, as
expected e.g. in SO(10), which allows simultaneously for massive
neutrinos\cite{susyrev}.

For  large values of $\tb$ the annihilation via pseudoscalar Higgs
exchange, being proportional to $(\tb)^2$, becomes dominant. The
annihilation cross section for a correct relic density requires
relatively light pseudoscalar Higgs masses, typically below or
around 500 GeV, which is consistent with the values given in Table
\ref{t1}. It should be noted that the annihilation is still in the
so-called bulk region, i.e. the regions not dominated by
co-annihilation or resonances, since the neutralino mass is far away
from the pseudoscalar Higgs mass resonance for the mSUGRA spectrum
and not close to any of the other sparticles, like stau or chargino,
as shown in Table \ref{t1}. If these latter Next-to Lightest
Supersymmetric Particles (NLSP) are almost degenerate with the LSP,
their number density, given by the Boltzmann factor, would  be high
enough to cause a fast annihilation in the early universe into taus
and charged W-bosons. The total annihilation rate, which is the sum
of the self-annihilation and co-annihilation rate, is fixed by the
observed relic density. Therefore a large co-annihilation rate
automatically implies a negligible self-annihilation rate. Since in
the present universe the NLSPs have decayed, only the
self-annihilation is operative now and would be practically zero in
case of strong co-annihilation. So it is fortunate for indirect DM
detection that the combination of EGRET data with the Higgs mass
limit results in a spectrum, for which the co-annihilation is
negligible.

\begin{figure}
 \begin{center}
  \includegraphics [width=0.4\textwidth,clip]{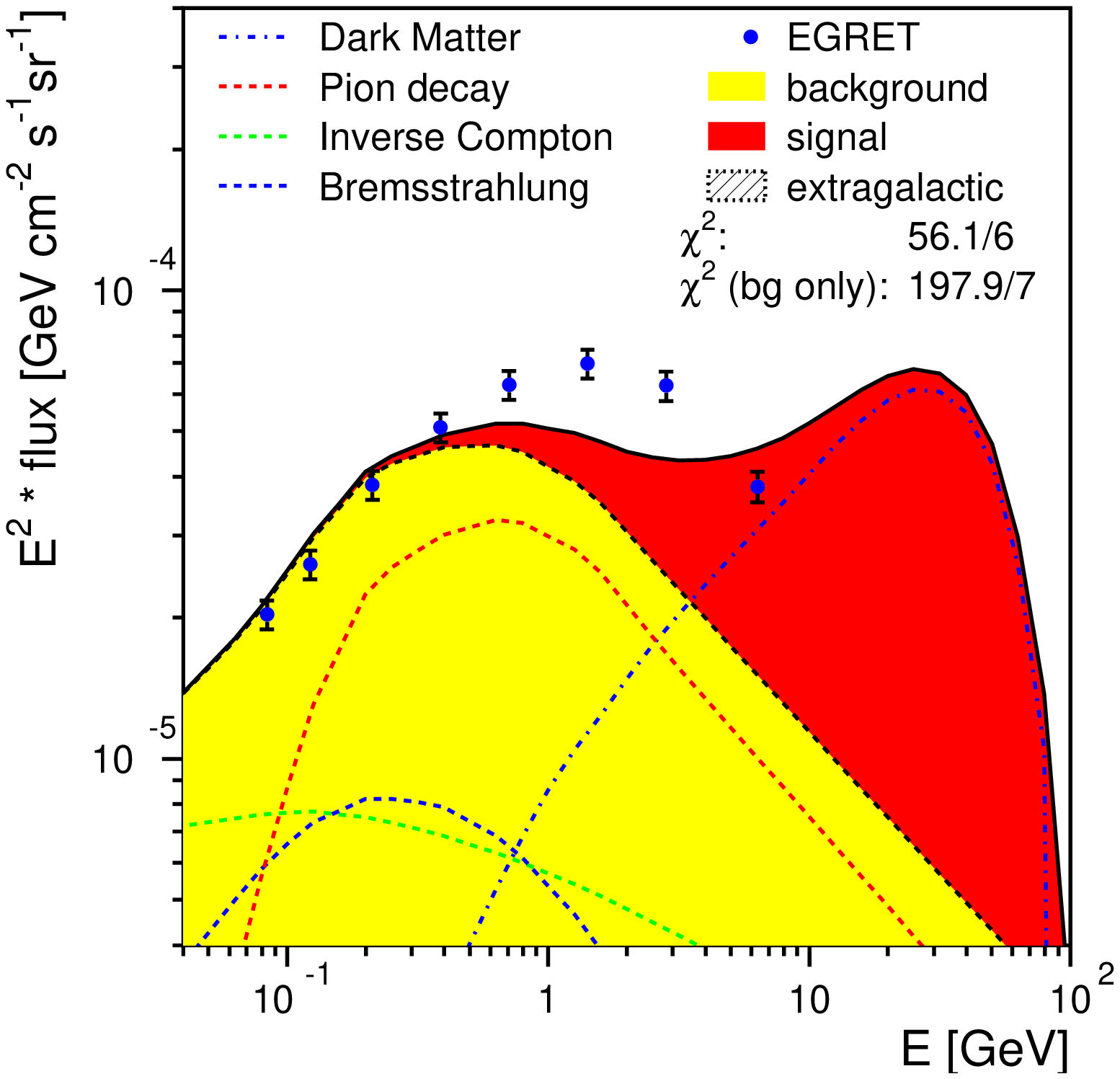}
  \includegraphics [width=0.4\textwidth,clip]{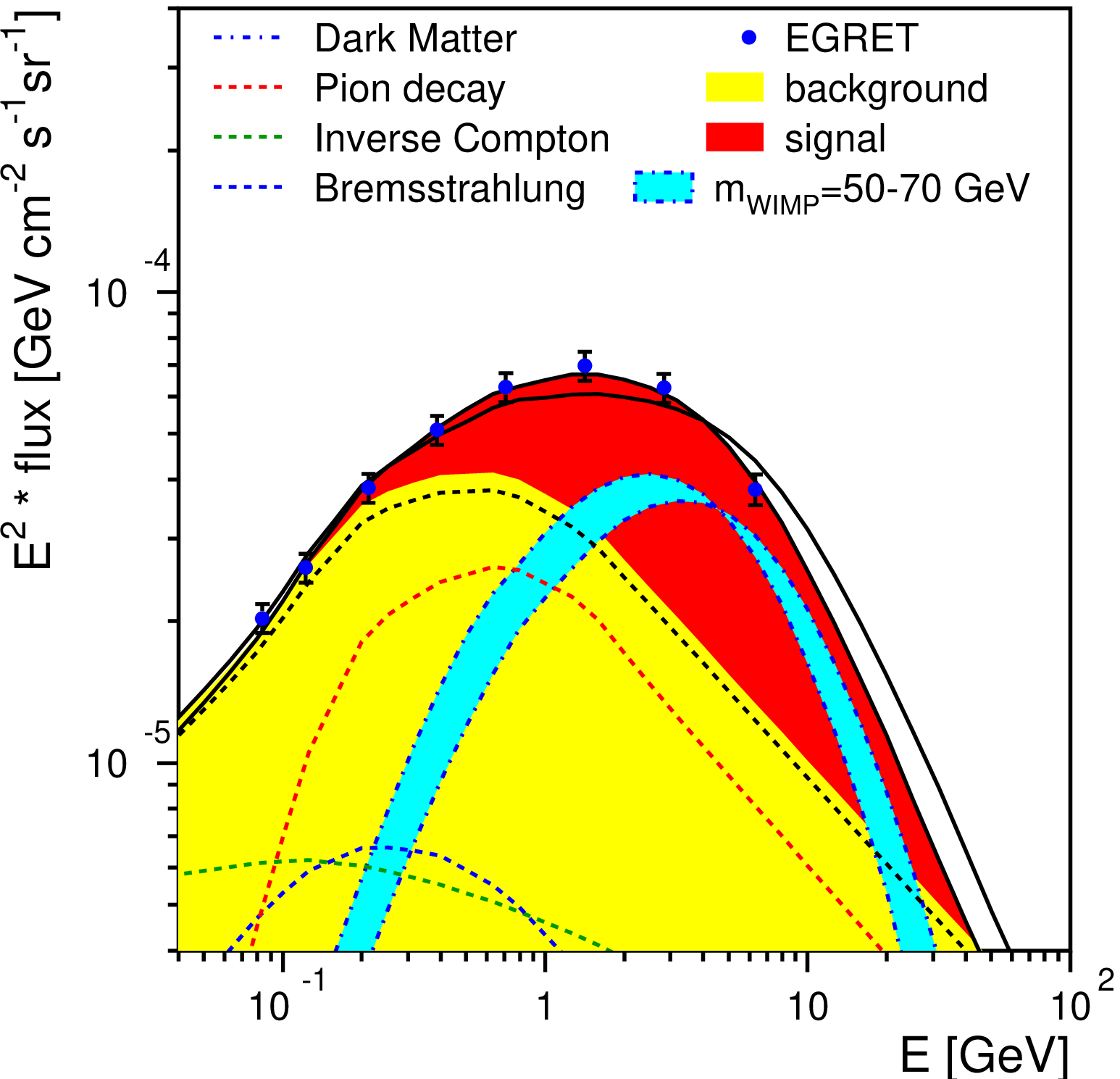}
  \caption[]{The EGRET gamma ray spectrum fitted with DM annihilation for
  ($\mzero =70$ GeV, $\mhalf =250$ GeV, $\tb=10$) (left) and ($\mzero =1400$ GeV,
  $\mhalf =175$ GeV, $\tb=51$) (right). In both cases the relic density corresponds to
  the WMAP value, but in the first case of low $m_0$ the annihilation into tau pairs dominates,
  while in the latter case the annihilation into $b$-quarks dominates.
  The first case is excluded by the EGRET data. On the right hand side the variation of
  the WIMP mass between 50 and 70 GeV ($m_{1/2}$ between 125 and 175 GeV) is shown as well
  (blue shaded area), which is the range allowed by the EGRET data with the conventional
  background (see Ref. \cite{us}).} \label{dsflux}
 \end{center}
\end{figure}

An independent check that the scalars should be heavy comes directly
from the EGRET data: if the scalars are light, the stau is usually
the lightest scalar, in which case the stau exchange in the
t-channel (left diagram of Fig. \ref{diagrams}) would be dominant,
thus leading to tau final states. The low decay multiplicity of tau
leptons leads to a much harder gamma ray spectrum from the hadronic
decays, which is excluded by the EGRET data, as shown in Fig.
\ref{dsflux}. This plot has been made for a neutralino mass of about
100 GeV in order to obey the Higgs limit with the low value of $m_0$
required for the stau exchange to be dominant. A neutralino of 50
GeV would still yield a maximum in the spectrum around 20 GeV in
disagreement with the EGRET data, which shows a maximum excess
around 2 GeV.

\begin{figure}
 \begin{center}
  \includegraphics [width=0.7\textwidth,clip]{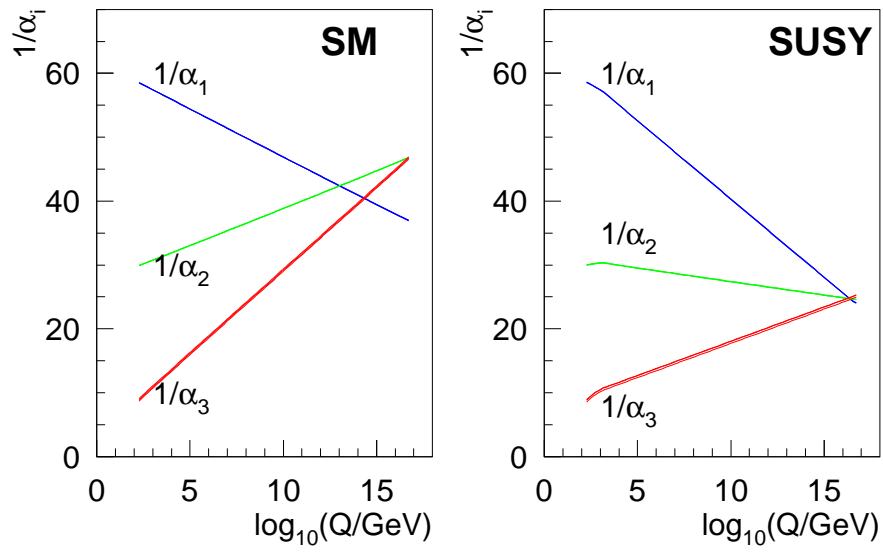}
  \caption[]{The running of the inverse of the gauge couplings in the SM (left) and in Supersymmetry with the SUSY mass spectrum from Table \ref{t1} (right).} \label{f5}
 \end{center}
\end{figure}

The SUSY spectrum of Table \ref{t1} yields excellent gauge
unification, as shown in Fig. \ref{f5}.  The used value of
$\alpha_s=0.122$ was taken from the ratio $R_l$ of the hadronic and
leptonic width of the $Z^0$ boson was taken, since the averaged LEP
value of 0.118 is the average of 0.115 from the hadronic cross
section $\sigma_h$ at LEP and 0.122 from $R_l$. However, the value
of 0.115 becomes 0.122 as well, if the luminosity at LEP is
normalized such that the number of neutrino generations is moved
from 2.98 to 3 \cite{bs}.  Note that in contrast to the earlier
evidence for gauge unification, where the SUSY mass scale had to be
taken as a free parameter \cite{unification}, there are now {\it no}
free parameters anymore since the initial starting points of the
running coupling constants are given by the electroweak precision
data from LEP and the change in the running from the SM to the MSSM
value is determined by the allowed masses in Fig. \ref{msugra} or
Table \ref{t1}. So one either gets unification or one does not get
it. Using the EGRET data and Higgs constraints one {\it gets}
unification in mSUGRA.
\section{Conclusion}

In our previous paper \cite{us} the observed excess of diffuse
Galactic gamma rays was shown to exhibit all  features of Dark
Matter Annihilation, especially the spatial distribution of the
excess was shown to trace the DM distribution, as proven by the fact
that one could reconstruct the peculiar shape of the rotation curve
of our Galaxy from the gamma ray excess. In this paper the DM
interpretation of the EGRET excess is compared with Supersymmetry
and it is shown that the minimal supersymmetric model with the
popular supergravity inspired symmetry breaking, gauge unification
and radiative electroweak symmetry breaking is in perfect agreement
with the EGRET excess. The mass spectrum of the gauginos is governed
by the neutralino mass corresponding to $\mhalf$ between 125 and 175
GeV, if the conventional background model is chosen. In case of a
model maximizing the background (optimized model, see Ref.
\cite{us}) values of $\mhalf$ up to 230 GeV are allowed. For the low
values of $\mhalf$ the scalar masses are constrained by the Higgs
mass and/or $\bsg$ to have $\mzero$ above 1 TeV.   The allowed mass
spectrum  is observable at the LHC. If confirmed, especially a
neutralino mass around 60 GeV, then this would prove that DM can
indeed be considered to be the supersymmetric partner of the Cosmic
Microwave Background, since the neutralino is almost a pure bino in
this case.

\newpage

\begin{table}
 \begin{center}
  \begin{tabular}{|c|c|}
   \hline
   Parameter & Value \\
   \hline
   $\mzero$ & 1500 GeV \\
   $\mhalf$ & 170 GeV \\
   $A_0$ & $0\cdot m_0$ \\
   $\tb$ & 52.2 \\
   sign $\mu$ & + \\
   \hline
   $\alpha_s(M_Z)$ & 0.122 \\
   $\alpha_{em}(M_Z)$ & 0.0078153697 \\
   $1/\alpha_{em}$ & 127.953 \\
   $\sin^2(\theta_W)_{\overline{MS}}$ & 0.2314 \\
   $m_t$ & 175 GeV\\
   $m_b$ & 4.214 GeV\\
  \hline
  \end{tabular}
  \begin{tabular}{|c|c|}
   \hline
   Particle & Mass [GeV] \\
   \hline
   $\tilde \chi^0_{1,2,3,4}$ & 64, 113, 194, 229 \\
   $\tilde \chi^\pm_{1,2},\tilde{g}$ & 110, 230, 516 \\
   $\tilde u_{1,2}=\tilde c_{1,2}$ & 1519, 1523 \\
   $\tilde d_{1,2}=\tilde s_{1,2}$ & 1522, 1524 \\
   $\tilde t_{1,2}$ & 906, 1046 \\
   $\tilde b_{1,2}$ & 1039, 1152 \\
   $\tilde e_{1,2}=\tilde \mu_{1,2}$ & 1497, 1499 \\
   $\tilde \tau_{1,2}$ & 1035, 1288 \\
   $\tilde \nu_e, \tilde \nu_\mu, \tilde \nu_\tau$ & 1495, 1495, 1286 \\
   $h,H,A,H^\pm$ & 115, 372, 372, 383 \\
   \hline
   Observable & Value \\
   \hline
   $Br(b\to X_s\gamma)$ & $3.02 \cdot 10^{-4}$ \\
   $\Delta a_\mu$ & $1.07\cdot 10^{-9}$ \\
   $\Omega h^2$ & 0.117 \\
   \hline
  \end{tabular}
  \caption[]{Typical mSUGRA parameters from the EGRET analysis and electroweak
  constraints.
  The corresponding mass spectrum of the SUSY particles and observables is shown on the right.} \label{t1}
 \end{center}
\end{table}

\end{document}